\documentclass[twocolumn,preprintnumbers,amsmath,amssymb,prb]{revtex4}

\usepackage{graphicx}
\usepackage{dcolumn}
\usepackage{bm}
\usepackage{amssymb}
\usepackage{epstopdf}
\usepackage{color}

\begin{document}

\title{Optically induced Kondo effect in semiconductor quantum wells}

\author{I. V. Iorsh$^{1,2}$}
\author{O. V. Kibis$^{2}$}\email{Oleg.Kibis(c)nstu.ru}
\affiliation{$^1$Department of Physics and Engineering, ITMO~ University, Saint-Petersburg, 197101, Russia}
\affiliation {$^2$Department of Applied and Theoretical Physics, Novosibirsk~State~Technical~University,
Karl~Marx~Avenue~20,~Novosibirsk~630073,~Russia}

\begin{abstract}
It is demonstrated theoretically that the circularly polarized irradiation of two-dimensional electron systems can induce the localized electron states which antiferromagnetically interact with conduction electrons, resulting in the Kondo effect. Conditions of experimental observation of the effect are discussed for semiconductor quantum wells.
\end{abstract}

\maketitle
\section{Introduction}
In 1964, Jun Kondo in his pioneering article~\cite{kondo1964resistance} suggested the physical mechanism responsible for the minimum of temperature dependence of the resistivity of noble divalent metals, which had remained a mystery for more than three decades~\cite{de1934electrical}. Within the developed theory, he showed that the antiferromagnetic interaction between the spins of conduction electrons and electrons localized on magnetic impurities leads to the $\log(T)$ corrections to the relaxation time of conduction electrons (the Kondo effect). The subsequent studies on the subject~\cite{wilson1975renormalization,Wiegmann_1981,Andrei_1983} demonstrated that physics of the Kondo effect is universal to describe the transformation of the ground state of various many-body systems in the broad range of energies. Particularly, the transformation is characterized by the single energy scale $T_K$ (the Kondo temperature) and can be effectively treated by the powerful methods of the renormalization group theory. Therefore, the Kondo problem is currently considered as an effective testing
ground to solve many challenging
many-body problems, including heavy-fermion materials, high-temperature
superconductors, etc~\cite{steglich1979superconductivity,andres19754,tsunetsugu1993phase,RevModPhys.56.755}.

The Kondo temperature is defined by the Coulomb repulsion of the impurity atoms, hybridization of the conduction and impurity electrons, and other condensed-matter parameters which are fixed in bulk materials but can be effectively tuned in nanostructures. Since the first observation of the tunable Kondo effect in such nanostructures as quantum dots~\cite{goldhaber1998kondo}, it attracts the enormous attention of research community~\cite{cronenwett1998tunable, iftikhar2018tunable,park2002coulomb,Borzenets_2020}. While the tuning of Kondo temperature in nanostructures is usually achieved by stationary fields (produced, e.g., by the gate voltage~\cite{iftikhar2018tunable}), it should be noted that all physical properties of them can be effectively controlled also by optical methods. Particularly, it has been demonstrated that the resonant laser driving of the impurity spins (such as quantum dot spins or single atom spins in the optical lattices) allows for the control over the onset and destruction of the Kondo resonance~\cite{Latta_2011,Haupt_2013, Tureci_2011, Sbierski_2013, Nakagawa_2015}. An alternative optical way of controlling the Kondo effect could be based on the modification of electronic properties by an off-resonant high-frequency electromagnetic field (the Floquet engineering~\cite{Basov_2017,Oka_2019}), which became the
established research area of modern physics and resulted in many fundamental effects in various nanostructures~\cite{Goldman_2014,Bukov_2015,Lindner_2011,Savenko_2012,Iorsh_2017,Kibis_2016,Kibis_2017,Kozin_2018,Kozin_2018_1,Rechtsman_2013,Wang_2013,Glazov_2014,Torres_2014,Sentef_2015,Sie_2015,Cavalleri_2020}.
Since the frequency of the off-resonant field lies
far from characteristic resonant frequencies of the electron system, the field cannot be
absorbed and only ``dresses'' electrons (dressing field), changing
their physical characteristics. Particularly, it was demonstrated recently that a high-frequency circularly polarized dressing field crucially modifies the interaction of two-dimensional (2D) electron systems with repulsive scatterers, inducing the attractive area in the core of a repulsive potential~\cite{Kibis_2019}. As a consequence, the light-induced electron states localized at repulsive scatterers appear~\cite{Kibis_2020,Iorsh_2020}. Since such localized electron states are immersed into the continuum of conduction electrons and interact with them antiferromagnetically, the Kondo effect can exist. The present article is dedicated to theoretical analysis of this optically induced effect for 2D electron gas in semiconductor quantum wells (QWs).

The article is organized as follows. In the second section, the model of Kondo effect based on the dressing field approach is developed. The third section is dedicated to the analysis of the Kondo resonance and the Kondo temperature in QWs. The last two sections contain conclusion and acknowledgements.

\section{Model}
\begin{figure}[h!]
\centering\includegraphics[width=0.8\columnwidth]{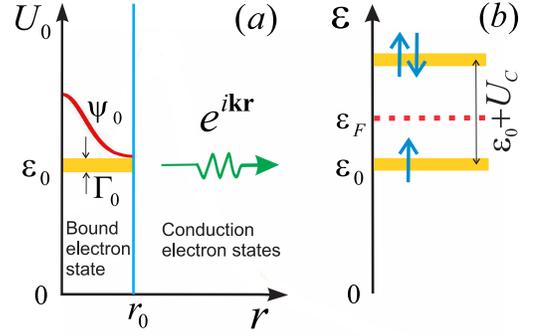}
\caption{Sketch of the system under consideration: (a) the optically induced potential (\ref{U0}) depicted by the vertical blue line, which confines the bound electron state (\ref{Elm}) marked by the horizontal yellow strip and separates it from the states of conduction electrons with the wave vectors $\mathbf{k}$ marked by the wave arrow; (b) the energy structure of singly-occupied ($\uparrow$) and doubly occupied ($\uparrow\downarrow$) electron states (\ref{Elm}) near the Fermi energy $\varepsilon_F$.} \label{fig:sketch}
\end{figure}

For definiteness, let us consider a semiconductor QW of the area $S$ in the plane $x,y$, which is filled by 2D gas of conduction electrons with the effective mass $m_e$ and irradiated by a circularly polarized electromagnetic wave incident normally to the $x,y$ plane. Then the behavior of a conduction electron near a scatterer with the repulsive potential $U(\mathbf{r})$ is described by the time-dependent Hamiltonian $\hat{\cal H}_e(t)=(\hat{\mathbf p}-e\mathbf{A}(t)/c)^2/2m_e+U(\mathbf{r})$, where $\hat{\mathbf p}$ is the plane momentum operator of conduction electron, $\mathbf{r}=(x,y)$ is the plane radius vector of the electron,
\begin{equation}\label{A}
\mathbf{A}(t)=(A_x,A_y)=[cE_0/\omega_0](\sin\omega_0
t,\,\cos\omega_0 t)
\end{equation}
is the vector potential of the wave, $E_{0}$ is the electric field
amplitude of the wave, and $\omega_0$ is the wave frequency. If the field frequency $\omega_0$ is high enough and lies far from characteristic resonant frequencies of the QW, this time-dependent Hamiltonian can be reduced to the effective stationary Hamiltonian, $\hat{\cal H}_0=\hat{\mathbf p}^2/2m_e+U_0(\mathbf{r})$, where
\begin{equation}\label{U0}
U_0(\mathbf{r})=\frac{1}{2\pi}\int_{-\pi}^{\pi}U\big(\mathbf{r}-\mathbf{r}_0(t)\big)\,d(\omega_0
t)
\end{equation}
is the repulsive potential modified by the incident field (dressed potential), $\mathbf{r}_0(t)=(-r_0\cos\omega_0 t,\,r_0\sin\omega_0 t)$ is the radius-vector describing the classical circular trajectory of a free electron in the circularly polarized field (\ref{A}), and $r_0={|e|E_0}/{m_e\omega^2_0}$ is the radius of the trajectory~\cite{Kibis_2019,Kibis_2020}. In the case of the short-range scatterers conventionally modelled by the repulsive delta potential,
\begin{equation}\label{delta}
U(\mathbf{r})=u_0\delta(\mathbf{r}),
\end{equation}
the corresponding dressed potential (\ref{U0}) reads~\cite{Kibis_2020}
\begin{equation}\label{U0}
U_0(\mathbf{r})=\frac{u_0\,\delta({r}-{r}_0)}{2\pi r_0}.
\end{equation}
Thus, the circularly polarized dressing field (\ref{A}) turns the repulsive
delta potential (\ref{delta}) into the
delta potential barrier of ring shape (\ref{U0}), which defines dynamics of an electron near the scatterer. As a consequence, the bound electron states which are
localized inside the area fenced by the ring-shape barrier
($0<r<r_0$) appear. Certainly,
such bound electron states are quasi-stationary since they can
decay via the tunnel transition through the potential barrier (\ref{U0}) into
the continuum of conduction electrons. As a consequence, the
energy broadening of the localized states appears. In the following, we will restrict the analysis by the ground localized state with the energy $\varepsilon_0$, the energy broadening $\Gamma_0$ and the wave function $\psi_0(r)$ (see Fig.~\ref{fig:sketch}). Assuming the repulsive delta potential to be strong enough ($\alpha=2\hbar^2/m_eu_0\ll1$), the solution of the Schr\"odinger problem with the stationary potential (\ref{U0}) can be written approximately as~\cite{Kibis_2020}
\begin{align}\label{Elm}
&\varepsilon_{0}=\frac{\hbar^2\xi^2_{0}}{2m_er^2_0},\,\,\,{\Gamma}_{0}=\frac{2\varepsilon_{0}\alpha^2}{N^3_0(\xi_{0})J_1(\xi_{0})},\nonumber\\
&\psi_{0}(r)=\frac{J_0\left({\xi_{0}r}/{r_0}\right)}{\sqrt{\pi}r_0J_{1}(\xi_{0})}
\theta(r_0-r),
\end{align}
where $J_m(\xi)$ and $N_m(\xi)$ are the Bessel functions of the
first and second kind, respectively, $\xi_{0}$ is the first zero
of the Bessel function $J_0(\xi)$, and $\theta(r)$ is the Heaviside function. It follows from the theory of the dressed potential approach~\cite{Kibis_2019} that the discussed model based on the dressed potential (\ref{U0}) is applicable if the dressing field frequency, $\omega_0$, much exceeds the characteristic frequency of the bound electron state, $\varepsilon_0/\hbar$.

Assuming the condition $\hbar\omega_0\gg\varepsilon_0$ to be satisfied, interaction between the localized electron state (\ref{Elm}) and the conduction electrons can be described by the Hamiltonian
\begin{eqnarray}
\hat{\cal H}&=&\sum_{\mathbf{k},\sigma} (\varepsilon_{\mathbf{k}}-\varepsilon_F)\hat{c}_{\mathbf{k}\sigma}^{\dagger}\hat{c}_{\mathbf{k}\sigma}+\sum_{\sigma}(\varepsilon_{0}-\varepsilon_F)\hat{d}_{\sigma}^{\dagger}\hat{d}_{\sigma}\nonumber\\ &+&U_C\hat{d}_{\uparrow}^{\dagger}\hat{d}_{\uparrow}\hat{d}_{\downarrow}^{\dagger}\hat{d}_{\downarrow}+\sum_{\mathbf{k},\sigma}T_{\mathbf{k}} \left[\hat{c}_{\mathbf{k},\sigma}^{\dagger}\hat{d}_{\sigma}+\mathrm{H.c.}\right], \label{eq:HT}
\end{eqnarray}
where $\varepsilon_{\mathbf{k}}=\hbar^2k^2/2m_e$ is the energy spectrum of conduction electrons, $\mathbf{k}$ is the electron wave vector, $\varepsilon_F$ is the Fermi energy of conduction electrons, $\sigma=\uparrow,\downarrow$ is the spin quantum number,
$\hat{c}_{\mathbf{k},\sigma}^{\dagger}$($\hat{c}_{\mathbf{k},\sigma}$) are the production (annihilation) operators for conduction electron states,
$\hat{d}_{\sigma}^{\dagger}$($\hat{d}_{\sigma}$) are the production (annihilation) operators for the light-induced localized electron states (\ref{Elm}),
\begin{eqnarray}\label{U}
U_C=e^2\int_{S}d^2\mathbf{r}\int_{S}d^2\mathbf{r}^\prime\frac{|\psi_{0}({r})|^2|\psi_{0}({r^\prime})|^2}{|\mathbf{r}-\mathbf{r}^\prime|} =\frac{\gamma e^2}{\epsilon r_0},
\end{eqnarray}
is the Coulomb interaction energy of two electrons with opposite spins $\sigma=\uparrow,\downarrow$ in the state (\ref{Elm}), $\epsilon$ is the permittivity of QW, $\gamma\approx0.8$ is the numerical constant, and $T_{\mathbf{k}}$ is the tunneling matrix element connecting the localized electron state (\ref{Elm}) and the conduction electron state with the wave vector $\mathbf{k}$. Physically, the first term of the Hamiltonian (\ref{eq:HT}) describes the energy of conduction electrons, the second term describes the energy of the localized electron in the state (\ref{Elm}), the third therm describes the Coulomb energy shift of the double occupied state (\ref{Elm}), and the fourth term describes the tunnel interaction between the conduction electrons and the localized electrons. Assuming the tunneling to be weak enough, one can replace the matrix element $T_{\mathbf{k}}$ with its resonant value, $|T_{\mathbf{k}}|^2=\Gamma_0/\pi N_\varepsilon$, corresponding to the energy $\varepsilon_{\mathbf{k}}=\varepsilon_0$, where $N_\varepsilon=Sm_e/\pi\hbar^2$ is the density of conduction electron states (see, e.g., Appendix B in Ref.~\onlinecite{Kibis_2020}). If the localized and delocalized (conduction) electron states are decoupled from each other ($T_{\mathbf{k}}=0$), the localized eigenstates of the Hamiltonian (\ref{eq:HT}) correspond to the singly occupied state (\ref{Elm}) with the eigenenergy $\varepsilon_0-\varepsilon_F$ and the doubly occupied state (\ref{Elm}) with the eigenenergy $2(\varepsilon_0-\varepsilon_F)+U_C$, which are marked schematically in Fig.~1b by the symbols $\uparrow$ and $\uparrow\downarrow$, correspondingly. For completeness, it should be noted that the empty state (\ref{Elm}) is also the eigenstate of the considered Hamiltonian with the zero eigenenergy corresponding to the Fermi level. Since the Kondo effect originates due to the emergence of magnetic moment (spin) of a localized electron, it appears only if the singly occupied state is filled by an electron but the doubly occupied state is empty. Assuming the temperature to be zero, this corresponds to the case of  $\varepsilon_{0}-\varepsilon_F<0$ and $U_C>\varepsilon_F-\varepsilon_{0}$. Taking into account that the characteristic energies of the considered problem, $U_C$ and $\varepsilon_{0}$, depend differently on the irradiation amplitude $E_0$ and frequency $\omega_0$, the optically induced Kondo effect can exist only in the range of these irradiation parameters defined by the inequality
\begin{equation}\label{r0}
\sqrt{\frac{\hbar^2 \xi_{0}^2}{2m_e\varepsilon_F}}<r_0< \frac{\gamma e^2}{2\epsilon\varepsilon_F}+\sqrt{\frac{\hbar^2 \xi_{0}^2}{2m_e\varepsilon_F}+\left(\frac{\gamma e^2}{2\epsilon\varepsilon_F}\right)^2}.
\end{equation}

Mathematically, the Hamiltonian (\ref{eq:HT}) is identical to the famous Anderson Hamiltonian describing the microscopic mechanism for the magnetic moment formation in metals~\cite{anderson1961localized}. Therefore, one can apply the known Schrieffer-Wolff (SW) unitary transformation~\cite{coleman2015introduction} to turn the Hamiltonian (\ref{eq:HT}) into the Hamiltonian of the Kondo problem~\cite{Andrei_1983}. Assuming the condition (\ref{r0}) to be satisfied, we arrive at the Kondo Hamiltonian
\begin{equation}
\hat{\cal H}_{K}=\sum_{\mathbf{k}\sigma} (\varepsilon_{\mathbf{k}}-\varepsilon_F)\hat{c}_{\mathbf{k}\sigma}^{\dagger}\hat{c}_{\mathbf{k}\sigma}+J\bm{\sigma}(0)\cdot \mathbf{S}_0 -\frac{{V}}{2}\hat{\psi}_{\sigma}^{\dagger}(0)\hat{\psi}_{\sigma}(0), \label{eq:Heff}
\end{equation}
where $\hat{\psi}_{\sigma}(0)=\sum_{\mathbf{k}} \hat{c}_{\mathbf{k}\sigma}$ is the $\hat{\psi}(\mathbf{r})$ operator of conduction electrons at the repulsive delta potential ($\mathbf{r}=0$), ${\hat{\bm{\sigma}}}(0)=\hat{\psi}^{\dagger}(0){\hat{\bm{\sigma}}}\hat{\psi}(0)$ is the spin density of conduction electrons at $\mathbf{r}=0$,  $\mathbf{S}_0=\hat{d}^{\dagger}({{\hat{\bm{\sigma}}}}/{2})\hat{d}$ is the spin density of an electron in the localized state (\ref{Elm}), $\hat{\bm{\sigma}}=(\sigma_x,\sigma_y,\sigma_z)$ is the spin vector matrix, and the coupling coefficients $J$ and ${V}$ read
\begin{eqnarray}
J&=&\frac{\Gamma_{0}}{\pi N_\varepsilon}\left[\frac{1}{\varepsilon_{0}-\varepsilon_F+U_C}+\frac{1}{\varepsilon_F-\varepsilon_{0}}\right],\label{J}\\
{V}&=&-\frac{\Gamma_{0}}{2\pi N_\varepsilon}\left[\frac{1}{\varepsilon_{0}-\varepsilon_F+U_C}-\frac{1}{\varepsilon_F-\varepsilon_{0}}\right]. \label{V}
\end{eqnarray}
It should be noted that the denominators in Eqs.~\eqref{J}--\eqref{V} are the energy detunings between the singly occupied and empty states, $\varepsilon_F-\varepsilon_0$, and the singly and doubly occupied states, $\varepsilon_0-\varepsilon_F+U_C$. Since excitation of the empty (doubly occupied) state creates an electron (hole) in the Fermi sea, we will label the corresponding detunings as $D_e=\varepsilon_F-\varepsilon_0$ and $D_h=\varepsilon_0-\varepsilon_F+U_c$.

It should be stressed also that the SW transformation is only applicable to the case of weak coupling between the localized electron state (\ref{Elm}) and the conduction electrons as compared to the energy difference between the ground (singly occupied) and excited (empty and doubly occupied) localized states. As a consequence, the Hamiltonian (\ref{eq:Heff}) accurately describes the asymmetric Kondo problem under consideration for $\Gamma_{0}\ll [\varepsilon_F-\varepsilon_{0},\varepsilon_{0}-\varepsilon_F+U_C]$ and, therefore, the detunings $D_{e,h}$ are assumed to meet the condition $\Gamma_{0}\ll D_{e,h}$. Beyond the condition, the system enters the so-called mixed-valence regime, where the localized states (\ref{Elm}) with different occupancy become effectively coupled~\cite{riseborough2016mixed}. Although the mixed valence regime hosts a rich class of the interesting phase transitions, in the following we will focus exclusively at the Kondo regime, where the singly-occupied ground localized state is well separated from the excited states.

\section{Results and discussion}
\begin{figure}[h!]
\centering\includegraphics[width=.8\columnwidth]{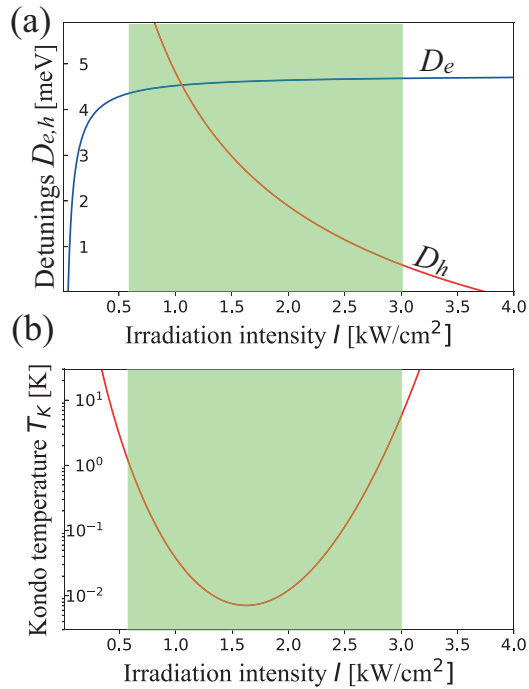}
\caption{Effect of the circularly polarized irradiation with the frequency $\omega_0/2\pi=200$~GHz and the intensity $I$ on (a) the hole and electron detunings $D_{h,e}$ and (b) the Kondo temperature in a GaAs-based QW filled by 2D electron gas with the Fermi energy $\varepsilon_F=5$~meV, energy broadening $\Gamma_0=0.1\varepsilon_0$  and the electron effective mass $m_e=0.067m_0$ ($m_0$ is the free electron mass) for the zero temperature. The green shadow areas mark the validity range of  the model, where the applicability conditions are satisfied for both the Kondo Hamiltonian ($\Gamma_{0}\ll D_e,D_h$) and the dressed potential approach ($\varepsilon_0\ll\hbar\omega_0$).} \label{fig:bands}
\end{figure}
For the particular case of 2D electrons in a GaAs-based QW, the dependence of the detunings $D_{e,h}$ on the irradiation is plotted in Fig.~2a. It should be noted that excitations of virtual electrons and holes should be considered within the whole conduction band of width $2D_0$, where $D_0\approx 1.5$~eV for GaAs. Since the typical Kondo temperature is essentially smaller than the bandwidth $D_0$, one needs to transform the initial high-energy Hamiltonian (\ref{eq:Heff}) to the low-energy range in order to find the Kondo temperature. Such a transformation can be performed  within the poor man's scaling renormalization approach~\cite{anderson1970poor,anderson1973kondo}, which was originally proposed by Anderson. Following Anderson, the higher energy excitations corresponding to the first order processes pictured in Fig.~3a can be integrated out from the Hamiltonian with using the SW transformation. Then the highest energy excitations, which are remained in the renormalized Hamiltonian, correspond to the second order processes pictured in Fig.~3b-c. It should be stressed that the renormalized Hamiltonian has the same structure as the initial one with the coupling constants depending on the renormalized (decreased) bandwidth $D<D_0$, where the renormalized coupling constant $J(D)$ diverges at some critical bandwidth $D=D_K$. This critical bandwidth defines the sought Kondo temperature, $T_K=D_K$, which, particularly, indicates the applicability limit of the perturbation theory~\cite{anderson1970poor,anderson1973kondo}.
\begin{figure}[h!]
\centering\includegraphics[width=1.\columnwidth]{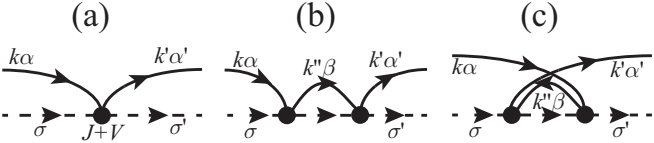}
\caption{The diagrams illustrating all possible first order processes (a) and the second order processes for electrons (b) and holes (c). The solid lines depict propagators of conduction electrons and holes, the dashed lines depict the localized spin propagator, whereas the symbols $\sigma$ and $\alpha$($\beta$) mark the spins of localized electrons and conduction electrons, respectively.} \label{fig:diagrams}
\end{figure}

The only difference of the considered system from the original Kondo problem~\cite{anderson1961localized} is the strong electron-hole asymmetry since the typical Fermi energy in GaAs-based QWs is $\varepsilon_F\ll D_0$. As a consequence, the hole process shown in Fig.~\ref{fig:diagrams}c cannot exists for $D>\varepsilon_F$ and only the second order process involving a virtual electron (see Fig.~\ref{fig:diagrams}b) should be taken into account. On the contrary, the both processes contribute to the effective coupling rescaling for the case of $D<\varepsilon_F$. Applying the known general solution of the asymmetric electron-hole Kondo problem~\cite{Zitko2016} to the considered system, the flow equations for the effective exchange constant $J^\prime(D)$ and the scalar potential $V^\prime(D)$ can be written as
\begin{align}
&\frac{1}{{\pi N_\varepsilon}}\frac{\partial {J}^\prime}{{\partial \ln(D_{0}/D)}}=[1+\theta(\varepsilon_F-D)]{{J^\prime}^2 -\theta(D-\varepsilon_F) {J^\prime}{V^\prime}},\nonumber\\
&\frac{2}{{\pi N_\varepsilon}}\frac{\partial {V}^\prime}{{\partial \ln(D_{0}/D)}}=-\theta(D-\varepsilon_F)\left[{3{J^\prime}^2+{V^\prime}^2}\right], \label{eq:vflow}
\end{align}
with the boundary conditions $J',V'|_{D=D_0}=J,V$. The solving of Eqs.~(\ref{eq:vflow}) should be performed in the two steps as follows. At the first step, we consider the interval $\varepsilon_F\leq D\leq D_0$. Within this interval, the two nonlinear differential equations~\eqref{eq:vflow} can be solved analytically (see, e.g., Ref.~\onlinecite{Zitko2016} for details) and result in the boundary condition
\begin{align}\label{boundary}
&J'(\varepsilon_F)=\\&=\frac{J}{\left[1+\frac{\pi N_{\varepsilon}}{2}\ln\frac{D_0}{\varepsilon_F}(J+V)\right]\left[1-\frac{\pi N_{\varepsilon}}{2}\ln\frac{D_0}{\varepsilon_F}(3J-V)\right]}.\nonumber
\end{align}
At the second step, we consider the interval $D\leq\varepsilon_F$. Within this interval, the scalar potential $V'$ is constant and the differential equation defining the effective exchange constant $J'$ does not depend on the scalar potential. Therefore, the system of two nonlinear differential equations~\eqref{eq:vflow} is reduced to the two independent differential equations. Solving them under the boundary condition (\ref{boundary}), one can find the effective exchange coupling constant $J'(D)$. Taking into account that the coupling constant diverges at the critical (Kondo) bandwidth $D=D_K$, we arrive the Kondo temperature
\begin{align}
T_K=\varepsilon_F\exp\left[-\frac{1}{2\pi N_{\varepsilon}J'(\varepsilon_F)}\right]. \label{eq:Tk}
\end{align}
Certainly, Eq.~(\ref{eq:Tk}) turns into the known expression for the Kondo temperature $T_K=D_0\exp[-1/2J\pi N_\varepsilon]$ corresponding to the symmetric Kondo problem~\cite{anderson1970poor} if $\varepsilon_F=D_0$ (the particular case of the half-filled band). The dependence of the Kondo temperature on the irradiation is plotted in Fig.~2b, where the Kondo temperature is found to be of several Kelvin. In the theory developed above, the field frequency, $\omega_0$, was assumed to satisfy the high-frequency condition $\omega_0\tau\gg1$, where $\tau$ is the mean free time of conduction electrons in the QW. To satisfy this condition for modern QWs and keep the irradiation intensity $I$ to be reasonable, we chose the field frequency for our calculations (see Fig.~2) near the upper limit of microwave range, $\omega_0/2\pi=200$~GHz, which can be easily realized in experiments. It should be noted also that the present theory is developed for the case of symmetric QW, although effects in asymmetric QWs are also studied actively (see, e.g., Refs.~\onlinecite{Stavrou_2001,Arulmozhi_2020}). In such asymmetric QWs, particularly, there is the Rashba spin-orbit coupling which can lead to the exponential increase of the Kondo temperature~\cite{Wong_2016}.

To observe the discussed effect experimentally, the known contribution of the Kondo resonance to the electron mean free time~\cite{kondo1964resistance},
\begin{align}\label{tau}
1/\tau\sim J^4\left[\frac{1}{\pi N_{\varepsilon}J}+ 2\ln \frac{D_0}{T}\right]^2,
\end{align}
can be used. Indeed, the found Kondo temperature (\ref{eq:Tk}) corresponds to the minimum of the contribution (\ref{tau}) as a function of the temperature $T$. Since all electron transport phenomena depend on the electron mean free time $\tau$, this minimum can be detected in various transport experiments (e.g., conductivity measurements). In order to exclude effects arisen from the irradiation-induced heating of electron gas, the difference scheme based on using both a circularly polarized field and a linearly polarized one can be applied. Indeed, the heating does not depend on the field polarization, whereas the electron states bound at repulsive scatterers --- and the related Kondo effect, respectively --- can be induced only by a circularly polarized field~\cite{Kibis_2019}.

\section{Conclusion}
We showed within the Floquet theory that a circularly polarized electromagnetic field irradiating a two-dimensional electron system can induce the localized electron states which antiferromagnetically interact with conduction electrons. As a consequence, the Kondo effect appears. For semiconductor quantum wells irradiated by a microwave electromagnetic wave of the intensity $\sim$kW/cm$^2$, the Kondo temperature is found to be of several Kelvin and, therefore, the effect can be detected in state-of-the-art transport measurements.

\section{Acknowledgements}
The reported study was funded by the Russian Science Foundation (project 20-12-00001).

\end{document}